\documentclass[useAMS,usenatbib,usegraphicx,fleqn]{mn2e}

\usepackage{amsmath}
\usepackage{subfigure}
\usepackage{graphicx}
\newcommand{\et}{{\it et al.}}

\newcommand{\simless}{\mathbin{\lower 3pt\hbox
     {$\rlap{\raise 5pt\hbox{$\char'074$}}\mathchar"7218$}}} %< or of order
\newcommand{\simgreat}{\mathbin{\lower 3pt\hbox
     {$\rlap{\raise 5pt\hbox{$\char'076$}}\mathchar"7218$}}} %> or of order

\title[Non-Equilibrium Populations of Hydrogen in High-Redshift Galaxies]{Non-Equilibrium Populations of Hydrogen in High-Redshift Galaxies}
\author[Brian B. Pomerantz, Kayla Redmond, and Vladimir Strelnitski]
{Brian B. Pomerantz$^{1,3}$, Kayla Redmond$^{2,3}$, and Vladimir Strelnitski$^{3}$\\
$^{1}$Cornell University, Ithaca, NY 14853\\
$^{2}$University of North Carolina-Ashville, 1 University Heights, Asheville, NC 28804\\
$^{2}$Maria Mitchell Observatory, 4 Vestal St, Nantucket, MA 02554, USA}

\numberwithin{equation}{section}
\numberwithin{table}{section}
\numberwithin{figure}{section}

\begin{document}

%\date{Accepted 20** Month **. Received 20** Month **; in original form 20** Month **}
\date{}
\pagerange{\pageref{firstpage}--\pageref{lastpage}} %\pubyear{20**}

\maketitle

\label{firstpage}

\begin{abstract}
We investigate the possibility of maser amplification in hydrogen recombination lines from the galaxies of first generation, at $z\la 30$. Combining analytical and computational approaches, we show that the transitions between the hydrogen Rydberg energy levels induced by the radiation from the ionizing star and by the (warmer than currently) Cosmic Microwave Background can produce noticeable differences in the population distribution, as compared with previous computations for contemporary H$^+$ regions, most of which ignored the processes induced by the ionizing star's radiation. In particular, the low ($n\la 30$) $\alpha$-transitions show an increased tendency toward population inversion, when ionization of the H$^+$ region is caused by a very hot star at high redshift. The resulting maser/laser amplification can increase the brightness of the emitted lines and make them detectable. However, the limiting effects of maser saturation will probably not allow maser gains to exceed one or two orders of magnitude.
\end{abstract}


\section{Introduction}

Detection of the first-generation galaxies from redshifts $z\la 30$ is one of the most important and most challenging goals of modern cosmology. Of particular interest is the possibility to detect any spectral lines from these galaxies, because such an observation would provide the key parameter of redshift. The prospects of detecting high-frequency hydrogen recombination lines (such as the low members of Lyman or Balmer series) from the ionized regions and the 21-cm line from the surrounding neutral gas have been intensively discussed (\textit{e.g.} Pritchard \& Loeb 2012).

Spaans \& Norman (1997) investigated the possibility of detecting high-$n$ hydrogen recombination lines amplified by the maser mechanism from the inhomogeneities of the expanding Universe at the epoch of recombination ($z\sim 1000$). They also mentioned the possibility of detectable masing $n\sim 120$ lines from the first galaxies, at $z\approx 30$, which, after redshift, fall into the low-frequency ($\sim 80\,$MHz) radio domain. In a recent study, Rule \et\ (2013) demonstrated that finding spontaneous emission in the high-$n$ hydrogen recombination lines in a blind search will probably be a challenge even for the most powerful modern facilities. Rule \et\ suggest that masing in some of these lines may help in making them detectable.

For maser amplification in a spectral line to occur, the populations of the corresponding energy levels must be inverted. The theoretical possibility of population inversion for the Rydberg levels of recombining hydrogen in the galactic H$^+$ regions first appeared in the calculations by Baker \& Menzel (1938), but it was overlooked by them (see Strelnitski 2012). The inversion was confirmed in many studies afterwards, starting with the important paper by Goldberg (1966) who pointed out that population inversion was implicitly present in the analytical results of Seaton (1964) and that it can increase considerably the intensities of high-$n$ radio recombination lines from H$^+$ regions. Later, the theoretical possibility of population inversion across large intervals of $n$, depending on the gas density and, less so, on its temperature, was demonstrated and studied by several investigators [\textit{e.g.} Salem \& Brocklehurst (1979), Walmsley (1990); Story \& Hummer (1995); Strelnitski \et\ (1996)].

From their analysis of the dominant reactions controlling the level populations, Baker \& Menzel (1938) came to the conclusion that all the processes induced by the radiation of the ionizing star, except for the photo-ionization from the ground level, are insignificant and can be ignored in the calculations.  This approach has dominated most of the subsequent comprehensive studies, including the most ambitious computations by Brocklehurst (1970) and by Storey \& Hummer (1995). In these and most other computations the level populations were presented as a function of only two parameters --- the electron temperature and density. The star's parameters (such as temperature, luminosity, distance from the region in question) were absent. The only external radiation that usually was taken into account was the black-body cosmic microwave background (CMB) with the contemporary temperature of $\approx 3\,$K. Note that Spaans \&\ Norman's (1997) numerical estimates of possible maser effects at the epoch of re-ionization were based on Storey \& Hummer's (1995) computations ignoring the radiation from the ionizing star.

Even if the ``$3\,$K CMB; no star'' approach can represent adequately the conditions in most of the H$^+$ regions of our own and other nearby galaxies, it may not be sufficient for understanding the situation in the galaxies of first generation. First of all, the CMB temperature was then considerably higher --- up to $\sim 100\,$K at $z\sim 30$. Not less importantly, the metal-poor ``Population III'' stars that produced the bulk of ionization in the early galaxies and beyond (causing the whole phenomenon of re-ionization of the Universe) were special --- very massive ($\ga 100 M_\odot$) and very hot ($\approx 100,000\,$K) (\textit{e.g.} Haiman \&\ Loeb 1997). Transitions induced by the radiation of such stars might have competed effectively with the spontaneous radiative processes and collisional processes in controlling hydrogen level populations.

The goal of the present study is to investigate the effects of the radiation of hot, massive stars and of a warmer CMB on the distribution of hydrogen level populations, including possible population inversion, at the epoch of first galaxy formation and to determine principal limitations of the strength of maser emission in hydrogen recombination lines from these galaxies. In Section~2 we provide general arguments in favor of the hypothesis that the effects caused by the star and CMB radiation may be significant. The computational procedure we used to check this assumption by numerical simulation is described in Section~3 and the results of the simulations --- in Section~4. Section~5 is devoted to an analysis of possible regimes of maser amplification in galaxies and to natural limitations of maser amplification. Section 6 contains some concluding remarks.

\section{Relevance of the Star and the CMB}
\subsection{The Star}
\label{sec:star}

In order to demonstrate the potential importance of radiative processes induced by the star for the distribution of level populations, we compare the rate of population supply by the transitions from the (most populated) ground level induced by the radiation of the star to the rates of two other processes that are considered to be of a primary importance: the radiative recombination and the spontaneous cascade from higher levels. We consider Menzel's Case B (an infinitely high optical depth in all the Lyman lines), for which the effective ground level is the level $n=2$, and calculate the ratios $\xi_1$ and $\xi_2$ of the excitation rate of level $n$ induced by the star's radiation to the rates of the two reference processes:
\begin{equation}
\label{eq:PEXratios}
\begin{split}
&\xi_1 \equiv \frac{N_2B_{2,n}J_{2,n}}{N_e^2\alpha_n^r},\\
&\xi_2 \equiv N_2B_{2,n}J_{2,n}\left(\sum_{i=n+1}^\infty N_iA_{i,n}\right)^{-1}\;,
\end{split}
\end{equation}
where $N_i$ is the population of level $i$; $J_{i,j}$ is the intensity of radiation averaged over the frequencies within the absorption profile of the line $i\to j$ and over directions; $A_{i,n}$ and $B_{i,n}$ are the Einstein coefficients; $N_e$ is the electron density; and $\alpha_n^r$ is the coefficient of radiative recombination to level $n$. (Here and afterwards our notations in equations are similar to those of Burgess \& Summers (1976), with the exception that we deal with the average intensity instead of the energy density of radiation.)

The calculations were done for constant values of the temperature of the star (considered to radiate as a black body), $T_* = 3.0\times 10^4\,$K, and the electron temperature, $T_e = 10^4\,$K. The dilution factors $W$ were calculated by considering the ionized gas at the periphery of the Str\"{o}mgren sphere (where the bulk of the mass of an ionization-bounded H$^+$ region is located).  Using table 2.3 from Osterbrock \& Ferland (2006), the following empirical equation connecting $W$ with $T_*$ and $N_e$ was obtained:
\begin{equation}
\label{eq:W}
W = 2.63\times 10^{-14}\left(\frac{57.74}{N_e^2}e^{4.82\times 10^{-5}T_*}\right)^{-2/3}.
\end{equation}
For the crude estimates of relative rates of population supply, we assumed LTE populations for all the levels, except level $n=2$. The population of the latter was calculated by balancing the major processes of population supply and depletion:
\begin{equation}
\label{eq:N2}
\begin{split}
N_e^2\alpha^r_2 + \sum_{i=3}^\infty N_iA_{i,2} &= N_2\left(\int B_{2,\kappa}J(\nu)\>\rmn{d}\kappa \right. \\
&+ \left.\sum_{i=3}^\infty N_eq_{2,i}^e + \sum_{i=3}^\infty B_{2,i}J_{2,i}\right)\;,
\end{split}
\end{equation}
where $J(\nu)$ is the intensity of ionizing radiation from the star averaged over directions, $B_{2,\kappa}$ is the Einstein $B$ coefficient for the bound-free transition from level 2 to a virtual level $\kappa$ of the continuum, and $q_{2,i}^e$ is the electron impact rate for transition $2\to i$.

Tables \ref{tab:PEX/AR} and \ref{tab:PEX/NA} show the results of calculations of $\xi_1$ and $\xi_2$ for various levels and a range of electron densities expected in the H$^+$ regions of the galaxies of first generation (Wood \& Loeb 2000). It is seen that the values of $\xi_1$ and $\xi_2$ depend very weakly on the electron density and that for the levels with $n\la 20$ the excitation of level $n$ from the ground level by the direct star's radiation is comparable with other processes controlling the populations.

The decrease of $\xi_1$ and $\xi_2$ with the increasing $n$ means that  photo-excitation by direct stellar radiation must have a {\it cooling} effect on population distribution.  In other words, when the radiation of the star is taken into account, we expect the inversion for $\alpha$-transitions predicted by the calculations ignoring the star to decrease.
\begin{table}
\centering
\caption{Values of $\xi_1$ given by equation (\ref{eq:PEXratios}) for $T_* = 3.0\times 10^4\,$K, $T_e = 10^4\,$K, and $W$ and $N_2$ calculated using equations (\ref{eq:W}) and (\ref{eq:N2}) respectively.}
\begin{tabular}{c|ccc}
	 &\multicolumn{3}{c}{$N_e\:({\rm cm}^{-3})$} \\
$n$ & $10^2$ & $10^4$ & $10^6$  \\ \hline
5   & 4.14   & 4.43   & 4.50    \\
10  & 1.39   & 1.49   & 1.51    \\
20  & 0.758  & 0.812  & 0.824   \\
30  & 0.590  & 0.632  & 0.642   \\
40  & 0.508  & 0.544  & 0.553   \\
\end{tabular}
\label{tab:PEX/AR}
\end{table}
\begin{table}
\centering
\caption{Values of $\xi_2$ given by equation (\ref{eq:PEXratios}) for $T_* = 3.0\times 10^4\,$K, $T_e = 10^4\,$K, and $W$ and $N_2$ calculated using equations (\ref{eq:W}) and (\ref{eq:N2}) respectively.}
\begin{tabular}{c|ccc}
	 &\multicolumn{3}{c}{$N_e\:({\rm cm}^{-3})$} \\
$n$ & $10^2$ & $10^4$ & $10^6$  \\ \hline
5   & 0.922  & 0.987  & 1.00    \\
10  & 0.373  & 0.400  & 0.406   \\
20  & 0.174  & 0.187  & 0.190   \\
30  & 0.114  & 0.122  & 0.124   \\
40  & 0.0848 & 0.0908 & 0.0922  \\
\end{tabular}
\label{tab:PEX/NA}
\end{table}

It can be shown analytically (see Appendix A) that in the range of high enough star temperatures, the influence of the stellar radiation --- the rate of photo-excitation from level $n = 2$, and thus the cooling effect on population distribution --- decreases with the increasing stellar temperature. This effect happens because as $T_*$ increases, the increase of the  depopulation rate of the ground level $n =2$ through photo-ionization proceeds faster than the increase of the rate of transitions $2 \to n$. One can expect, therefore, that the calculated degree of inversion will be greater in the presence of a hotter star. It can also be demonstrated analytically that for low enough star temperatures the effect should reverse, because at low $T_*$ the control of the depletion of the ground level population passes from photoionization to collisional processes. However, numerical simulations (Section 4) show that this happens at $T_* \simless 10,000\,$K, and such stars are inefficient in creating H$^+$ regions.

Although the bulk of the gas in a quasi-spherical, ionization-bounded H$^+$ region is situated at the periphery of the region, and the equation \eqref{eq:W} gives a crude estimate of the dilution factor there, it is of some interest whether the distribution of the hydrogen level populations, and, in particular, the degree of the population inversion, depends on the dilution factor (the distance to the ionizing star). Analytical arguments presented in Appendix B show that this dependence must be weak. Therefore, we consider our results obtained with the use of equation \eqref{eq:W} -- both in this section and in Section 4 -- as reliable for all the ionized gas of a galaxy.

\subsection{The CMB}
\begin{figure}
\includegraphics[width=84mm]{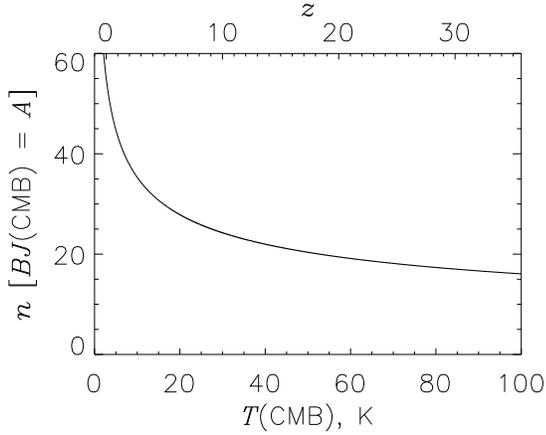}
%\vspace{3.5cm}
\caption{Dependence of the boundary transition for which $(BJ_{CMB}) \approx A$ on CMB temperature and redshift.}
\label{fig:1}
\end{figure}

In order to estimate the role of the CMB in the control of the level populations, we can compare the rate of transitions induced by the CMB with the rate of spontaneous decay of the levels, the latter dominating the population depletion of the  levels with $n\simless 100$ under the electron densities considered here ($N_e \simless 10^6\, \mathrm {cm^{-3}}$). With the use of the Planck equation for $J_{n,n'}$, the standard relation between the Einstein coefficients, and the well known equation for the frequency of the hydrogen $\alpha$-transitions, the condition of the CMB dominance,
\begin{equation}
\frac{B_{n,n'}J_{n,n'}}{A_{n,n'}} > 1\, ,
\end{equation}
reduces to
\begin{equation}
{\mathrm {Ry}}\, c\,[n^{-2} - (n+1)^{-2}]\la \frac{kT_{CMB}}{h}\,,
\end{equation}
where Ry is the Rydberg constant for hydrogen, $T_{\mathrm CMB}$ is the radiation brightness temperature, and the constants have their usual meaning. Figure \ref{fig:1} shows the dependence of the principal quantum number of the  boundary $\alpha$-transition for which $(BJ)_{CMB} \approx A$, as a function of the CMB temperature and the corresponding redshift $z$. It is seen that at the CMB temperatures of interest here ($\sim 20-100\,$K), a considerable influence of the CMB on the population distribution is expected for relatively high transitions, $n \ga 20$.

\section{Numerical Model and the Method of Solution}

In order to verify and specify the conclusions drawn in Section 2 using an analytical approach, we performed computer modeling of hydrogen Rydberg level populations under the assumed physical conditions  in the H$^+$ regions of the first generation galaxies. We follow Burgess \& Summers (1976) in the mathematical formulation of the problem. The populations are assumed to be in a steady state. The following transitions are accounted for: radiative excitation and de-excitation (spontaneous and stimulated), radiative ionization and recombination (spontaneous and stimulated), collisional excitation, de-excitation and ionization, and three-body recombination. The problem is reduced to the solution of a system of linear equations for level populations, $N_n$:
\begin{equation}
\begin{split}
&\sum_{n'>n} N_{n'}\left[A_{n',n} + B_{n',n}J_{n',n} + N_eq_{n',n}\right] \\
&+ \sum_{n''<n} N_{n''}\left[B_{n'',n}J_{n'',n} + N_eq_{n'',n}\right] \\
&- \sum_{n'>n} N_n\left[B_{n,n'}J_{n,n'} + N_eq_{n,n'}\right] \\
&- \sum_{n''<n} N_n\left[A_{n,n''} + B_{n,n''}J_{n,n''} + N_eq_{n,n''}\right] \\
&- N_n\int B_{n,\kappa}J_{n,\kappa}\>\rmn{d}\kappa + N_nN_eq_{n,\varepsilon} \\
&= N_e^2\alpha_n^r + N_e^3\alpha_n^3 + N_e^2\int B_{\kappa,n}J_{\kappa,n}\>\rmn{d}\kappa\,,
\end{split}
\label{eq:BS8}
\end{equation}
where $A_{n',n}$, $B_{n',n}J_{n',n}$, and $N_e q_{n',n}$ are the rate coefficients (dimensions [s$^{-1}]$) of spontaneous emission, stimulated emission and collisional transition, and $N_e^2\alpha_n^r$ and $N_e^3\alpha_n^3$ are the rates (dimensions [$\mathrm{cm^{-3}s^{-1}}]$) of radiative recombination and three-body collisional recombination, respectively.  The system consists of one instantiation of equation (\ref{eq:BS8}) for each level.

In a matrix form, the system of equations is: $C\vec{x} = \vec{b}$, where $C_{i,j\neq i}$ is the sum of the rate coefficients for all transitions to level $i$, $C_{i,i}$ is the rate coefficient for all transitions from level $i$ to any other level or the continuum, $\vec{x}_i$ is the unknown population of level $i$, and $\vec{b}_i$ is the sum of the rate coefficients for all transitions from the continuum to level $i$, as in equation (10) of Burgess \& Summers (1976).  We assumed the levels above $n=500$ to be distributed according to thermodynamic equilibrium.  The non-homogeneous system of equations was then solved using LU decomposition and back-substitution as described in Press \textit{et al.} (1986).

Finally, the parameter $\beta$ --- the ratio of the actual and the local thermodynamic equilibrium correction factors for stimulated emission, \textit{i.e.},
\begin{equation}
\beta_{n,n+1} = \frac{1-(b_{n+1}/b_n)\exp(-h\nu_0/kT_e)}{1-\exp(-h\nu_0/kT_e)},
\label{eq:betaDef}
\end{equation}
where $b_n$ is Menzel's coefficient of departure of the level $n$ population from its LTE value, and $\nu_0$ is the rest frequency of the $n \to (n+1)$ transition --- was calculated for each $\alpha$-transition. This parameter characterizes deviations of the population ratio from the LTE ratio. In particular, it indicates the presence and degree of population inversion: the transition is inverted when $\beta$ is negative, and the larger the magnitude of $\beta$, the higher the degrees of inversion (\textit{e.g.}, Strelnitski \et\ 1996).

\section{Results of Numerical Modeling}

\begin{figure}
\centering
\subfigure[]{
\includegraphics[width=84mm]{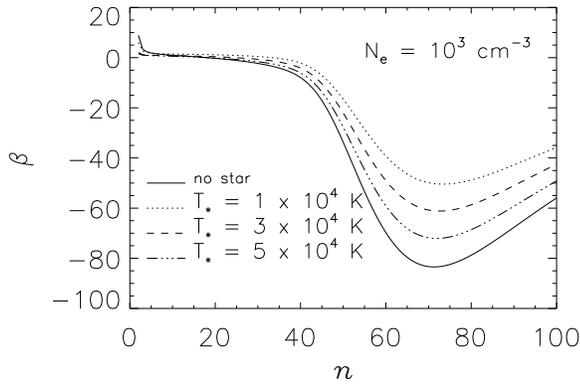}
%\vspace{3.5cm}
\label{fig:01}
}
\subfigure[]{
\includegraphics[width=84mm]{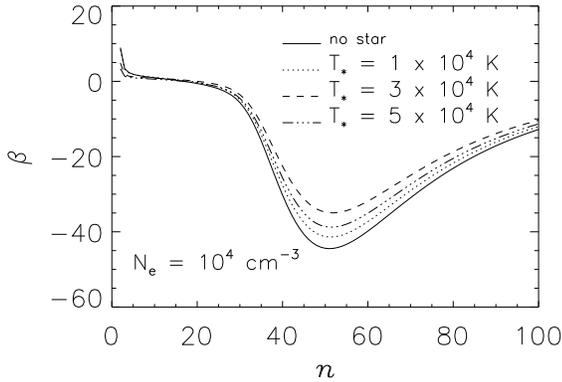}
%\vspace{3.5cm}
%\caption{Solid: $T_*=0\,$K, Dotted: $T_*=1.0\times 10^4\,$K, Dashed: $T_*=3.0\times 10^4\,$K, Dashed-Dotted: $T_*=5.0\times 10^4\,$K; at $N_e=10^4\,\rmn{cm^{-3}}$, $W=10^{-12}$, $T_{CMB}=3\,$K, $T_e=10^4\,$K, Case B}
\label{fig:02a}
}
\subfigure[]{
\includegraphics[width=84mm]{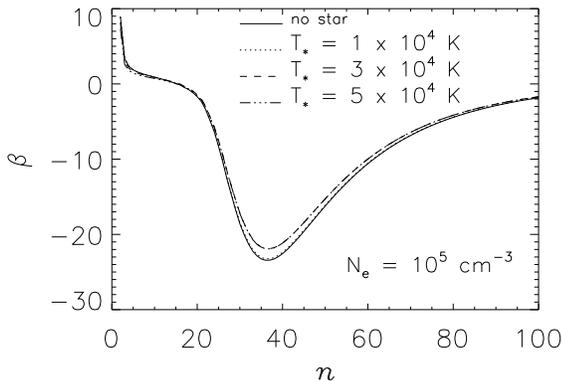}
%\vspace{3.5cm}
%\caption{Solid: $T_*=0\,$K, Dotted: $T_*=1.0\times 10^4\,$K, Dashed: $T_*=3.0\times 10^4\,$K, Dashed-Dotted: $T_*=5.0\times 10^4\,$K; at $N_e=10^4\,\rmn{cm^{-3}}$, $W=10^{-12}$, $T_{CMB}=3\,$K, $T_e=10^4\,$K, Case B}
\label{fig:02b}
}
\caption{Population inversion for various stellar temperatures and electron densities ($T_e=10^4\,$K, $T_{CMB}=3\,$K, $W=$ eq.(\ref{eq:W}), Case B).}
%\caption{Solid: $T_*=0\,$K, Dotted: $T_*=1.0\times 10^4\,$K, Dashed: $T_*=3.0\times 10^4\,$K, Dashed-Dotted: $T_*=5.0\times 10^4\,$K; at $N_e=10^3\,\rmn{cm^{-3}}$, $W=10^{-12}$, $T_{CMB}=3\,$K, $T_e=10^4\,$K, Case B}
\label{fig:1abc}
\end{figure}

\begin{figure}
\includegraphics[width=84mm]{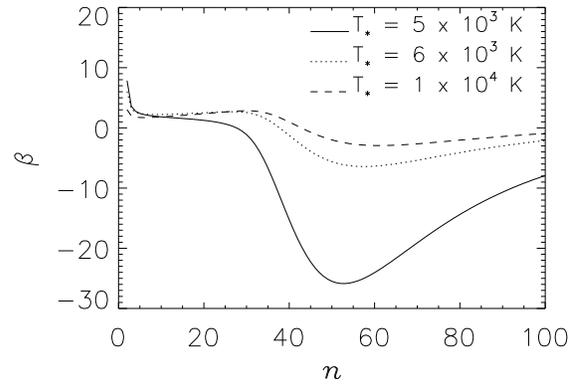}
\caption{Population inversion for low stellar temperatures ($T_e=10^4\,$K, $N_e=10^4\,\rmn{cm^{-3}}$, $T_{CMB}=3\,$K,  $W=$ eq.(\ref{eq:W}), Case B).}
\label{fig:11}
\end{figure}

\begin{figure}
\includegraphics[width=84mm]{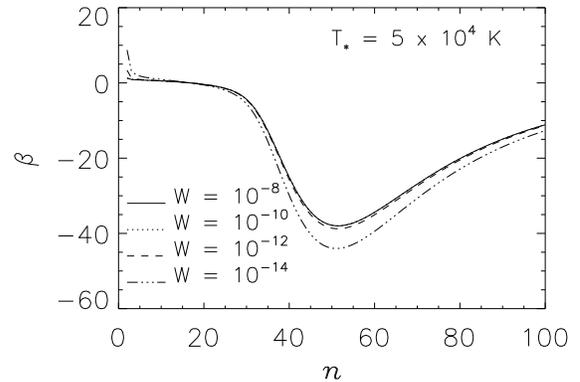}
%\caption{Solid: $W=10^{-8}$, Dotted: $W=10^{-10}$, Dashed: $W=10^{-12}$, Dashed-Dotted: $W=10^{-14}$; at $N_e=10^4\,\rmn{cm^{-3}}$, $T_*=5.0\times 10^4\,$K, $T_{CMB}=3\,$K, $T_e=10^4\,$K, Case B}
\label{fig:04}
\caption{Population inversion for various dilution factors ($N_e=10^4\,\rmn{cm^{-3}}$, $T_e=10^4\,$K, $T_{CMB}=3\,$K, Case B).}
%\caption{Solid: $W=10^{-8}$, Dotted: $W=10^{-10}$, Dashed: $W=10^{-12}$, Dashed-Dotted: $W=10^{-14}$; at $N_e=10^4\,\rmn{cm^{-3}}$, $T_*=5.0\times 10^4\,$K, $T_{CMB}=3\,$K, $T_e=10^4\,$K, Case B}
\label{fig:2ab}
\end{figure}

\begin{figure}
\centering
\subfigure[]{
\includegraphics[width=84mm]{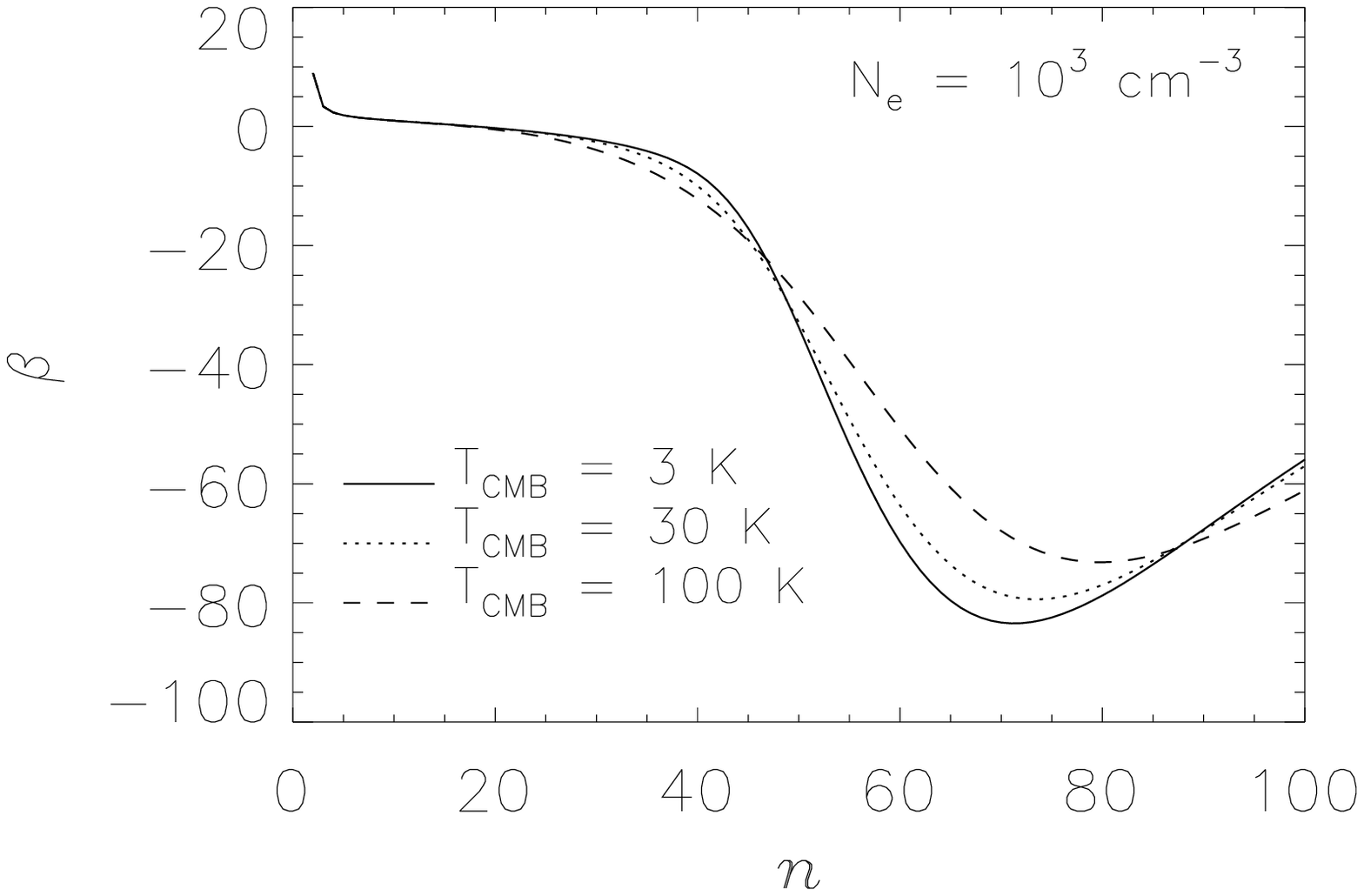}
%\caption{Solid: $T_{CMB}=3\,$K, Dotted: $T_{CMB}=30\,$K, Dashed: $T_{CMB}=100\,$K; at $N_e=10^3\,\rmn{cm^{-3}}$, $W=0$, $T_e=10^4\,$K, Case B}
\label{fig:05}
}

\subfigure[]{
\includegraphics[width=84mm]{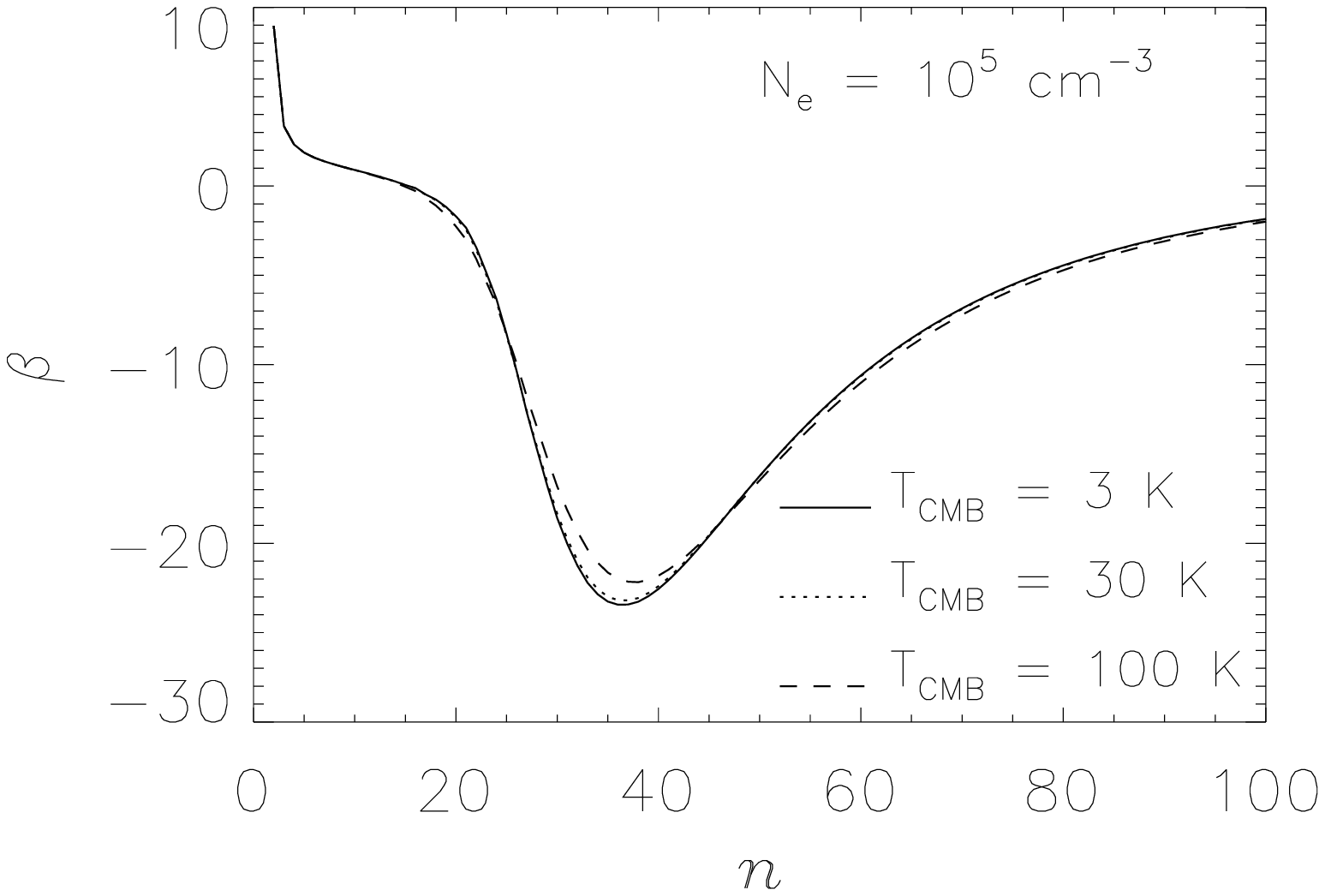}
\label{fig:06}
}
\caption{Population inversion for various CMB temperatures and electron densities ($T_e=10^4\,$K, $W=0$, Case B).}
\label{fig:3abc}
\end{figure}

\begin{figure}
\subfigure[]{
\includegraphics[width=84mm]{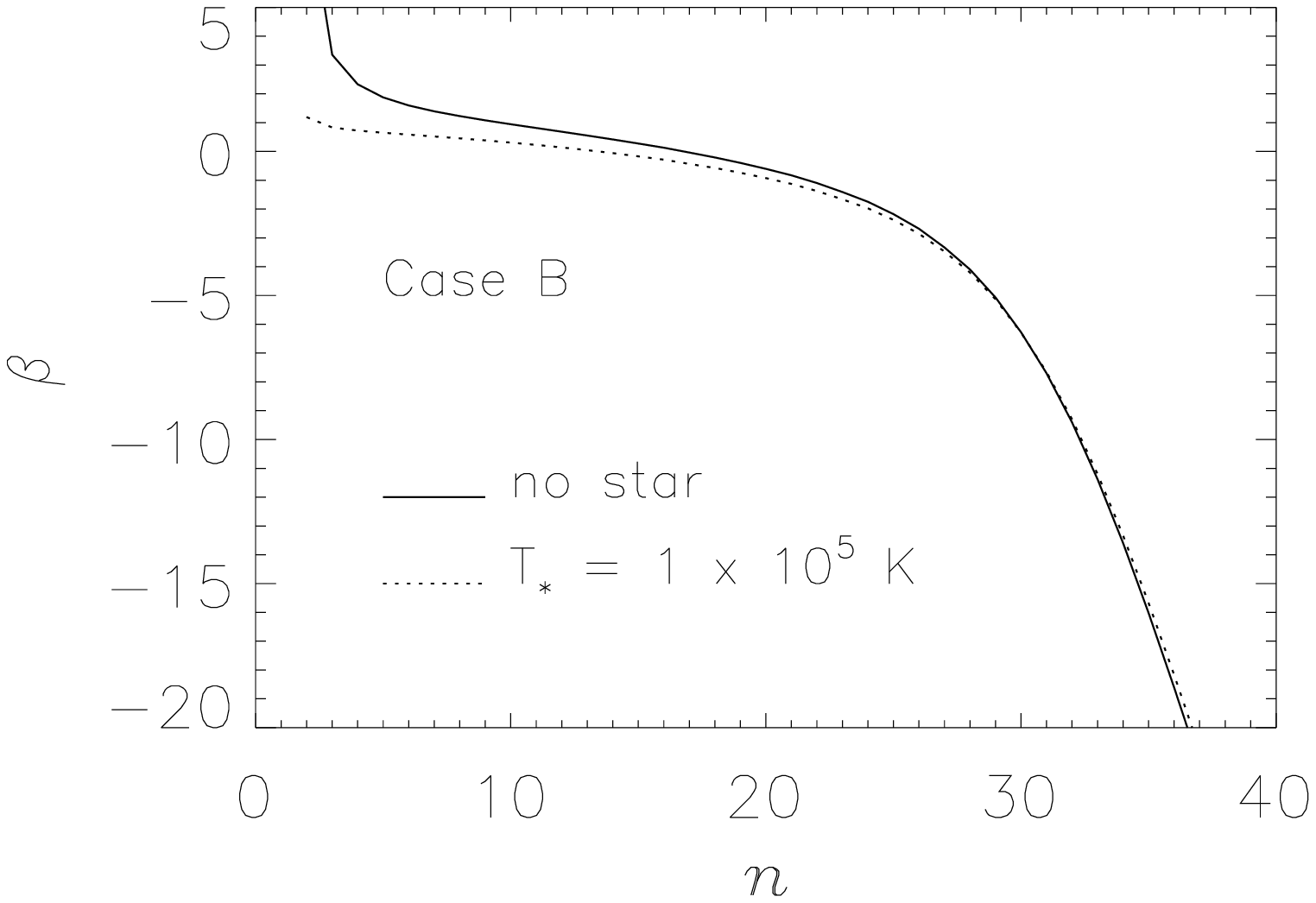}
%\caption{Solid: $T_*=0\,$K, Dotted: $T_*=1.0\times 10^5\,$K; at $N_e=10^4\,\rmn{cm^{-3}}$, $W=$ eq: \ref{eq:SSW}, $T_{CMB}=30\,$K, $T_e=10^4\,$K, Case B}
\label{fig:10}
}
\subfigure[]{
\includegraphics[width=84mm]{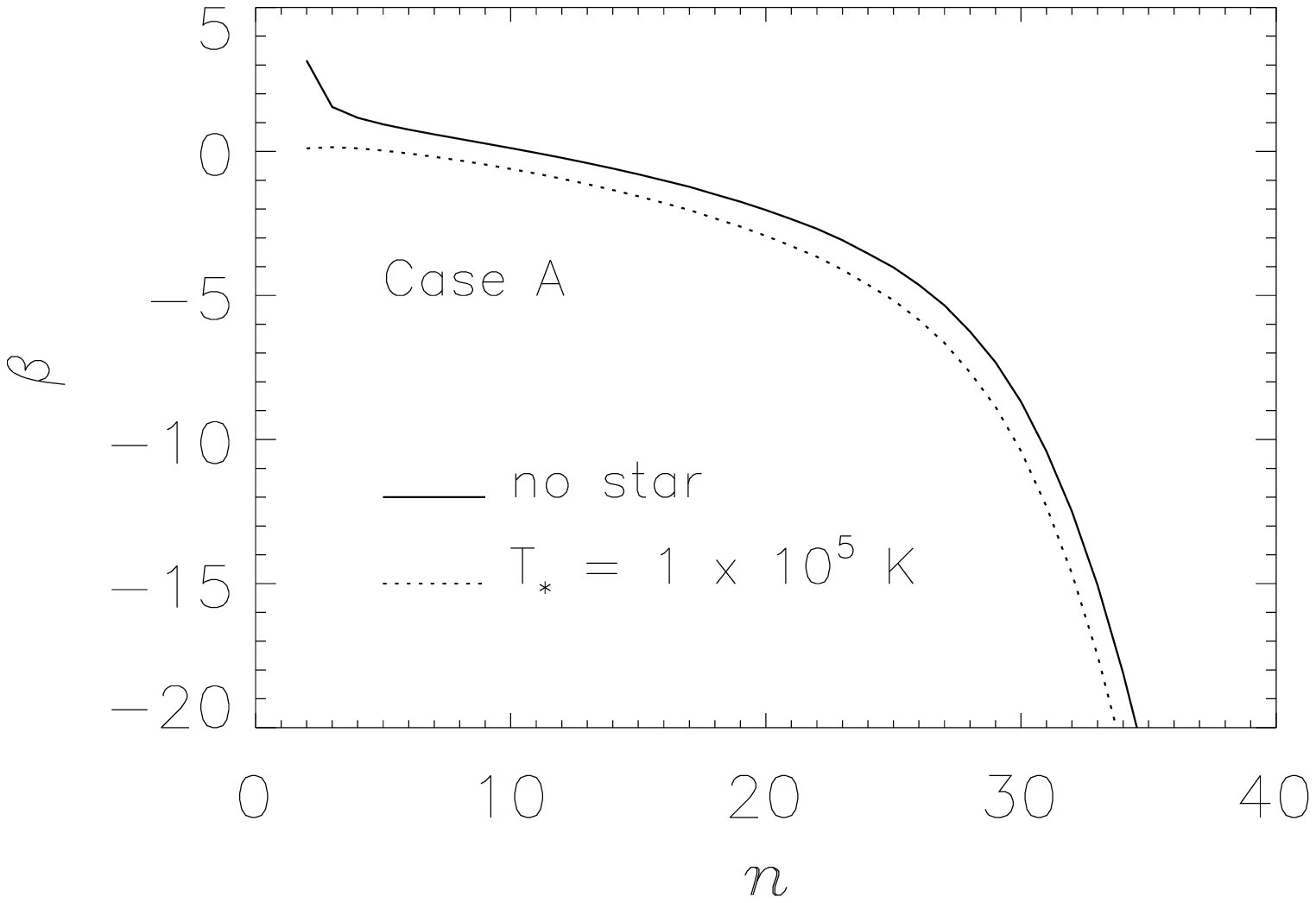}
\label{fig:10b}
}
\caption{Comparison of population inversion at low $n$ between the cases of ``no star'' and a hot ionizing star ($T_e=10^4\,$K, $N_e=10^4\,\rmn{cm^{-3}}$, $W=$ eq.(\ref{eq:W}), $T_{CMB}=30\,$K).}
\label{fig:5ab}
\end{figure}

The results of computations are presented in Figures 4.1 - 4.5.

Figure \ref{fig:1abc} demonstrates the dependence of parameter $\beta$ on the principal quantum number of an $\alpha$-transition for various electron densities and star temperatures. The general behavior of $\beta(n)$ is similar to that obtained in previous computations of hydrogen level populations (compare, for example, with Fig. 4a in Strelnitski et al. (1996)]. In particular, there is a broad minimum of negative $\beta$ (maximum of inversion), with the quantum number corresponding to the highest inversion decreasing when the electron density increases. Fig. \ref{fig:1abc} demonstrates that even for contemporary HII regions ($T_{CMB} \approx 3\,$K) the inclusion of the processes of population transfer induced by the radiation of the star in the computations may result in significant changes of the degree of inversion. The scale of the effect depends on the electron density, decreasing as $N_e$ increases.

Fig. \ref{fig:1abc} also shows that, as analytically predicted in section 2.2, the degree of inversion tends to increase for increasing stellar temperatures --- the ``no star'' case having the largest degree of inversion. This tendency holds for the whole range of stellar temperatures interesting for the present study, $T_* \ga 1\times 10^4\,$K. For lower $T_*$, the tendency reverses, because of the decreasing role of depopulation of the effective ground level $n=2$ by photodissociation, as compared with collisional depopulation. This reversal is illustrated in Fig. \ref{fig:11} for $T_* \la 1\times 10^4\,$K: in this temperature domain, with the decrease of $T_*$ the solution tends to that of ``no star.''

As argued in section \ref{sec:star} and Appendix B, variations in the dilution factor don't have a significant effect on the population distribution, which justifies the use of equation (2.2). The weakness of the solution dependence on $W$ is confirmed by the numerical simulation presented in Fig. 4.3. With the values of the parameters used in this simulation, the dilution factor determined by equation (2.2) is $W = 7.6\times 10^{-11}$. The curves, corresponding to $W$ two orders of magnitude lower and two orders of magnitude higher are practically the same.

Figure \ref{fig:3abc} illustrates the influence of the CMB on the level populations at various redshifts (various $T_{CMB}$). For more clarity, the population transfer processes stimulated by the radiation of the star were excluded in these numerical models ($W = 0$). The figure shows that at low densities (Fig.~4.4a) a significant increase in CMB temperature (between $T_{CMB} \approx 3\,$K and $100\,$K, i.e. the increase of $z$ between $\approx 0$ and 30) can cause a noticeable change in the population distribution. As anticipated (see Section 2), only transitions with $n \ga 20$ are affected. The most remarkable feature is the difference in the ``sign'' of the effect between the transitions around the maximum of inversion ($n \sim 50$ -- 80) and those with lower and higher $n$'s: the increase in CMB temperature causes a decrease in the degree of inversion for the former and a slight increase for the latter.  These effects, however, decrease for higher electron densities, as seen in Fig.~4.4b.

The expected combined effects produced on low-$n$ transitions by the CMB and the radiation of relatively populous high-mass, high-temperature Population III stars in the first galaxies at $z\sim 10$ ($T_{CMB} \sim 30\,$K) are illustrated in Fig. 4.5. Although the increase of the degree of inversion is not large, for some of these transitions the effect may mean a jump from the state of no inversion to the state of inversion and thus a principal possibility of maser amplification.

\section{Maser Amplification in Galaxies}
\subsection{Regimes of Maser Amplification}

A specific feature of radio recombination lines from H$^+$ regions, especially important for lower frequency lines, is that they are formed together with the free-free continuum. Therefore the emission and absorption processes in both the line and the continuum must be included in the radiative transfer equation. The major effects of stimulated emission in radio recombination lines can be demonstrated with the simple model of a steady-state, homogeneous, plane-parallel ionized medium of finite optical thickness (hereafter ``the cloud''). The solution of the radiative transfer equation for such a cloud can be given the following useful form (Shaver 1978):
\begin{equation}
\label{eq:DelS}
\begin{split}
\Delta S_L \equiv S_L - S_C = B_{\nu}(T_e)\, &\omega \biggl [ \eta\, (1 - {\rm e}^{-(\tau_L + \tau_C)}) - (1 - {\rm e}^{-\tau_c})\biggr ]\\
&+\, S_0\,{\rm e}^{-\tau_C}({\rm e}^{-\tau_L} -1)\;,
\end{split}
\end{equation}
where $S_L$ and $S_C$ are the observed flux densities at the central frequency of the line and in the adjacent continuum, $B_{\nu}(T_e)$ is the Planck function, $\omega$ is the solid angle subtended by the cloud as seen from Earth,  $S_0$ is the (unattenuated) flux density due to a background source of continuum (which is supposed to subtend a solid angle $\le \omega$); $\tau_C$ and $\tau_L$ are the total optical depths of the cloud due to the continuum and line opacities, respectively. The parameter $\eta$, introduced by Goldberg (1966), is very nearly equal to the ratio of the line-plus-continuum opacities in and out of LTE:
\begin{equation}
\label{eq:eta}
\eta = \frac{k_C + k_L^*b_2}{k_C + k_L^* b_1 \beta_{12}} = \frac{\tau_C +\tau_L^*b_2}{\tau_C + \tau_L^* b_1 \beta_{12}}\;,
\end{equation}
where $k_L$, $k_C$ are the line and continuum absorption coefficients, and asterisks denote LTE. The non-LTE line opacity and optical depth are obtained by multiplying the corresponding LTE values by $b_1\beta_{12}$.

Three asymptotic cases of the line amplification can be extracted from  equation~(\ref{eq:DelS}):

{\it (1) Low opacities, no background continuum} (Goldberg 1966). Assuming $\tau_L, \tau_C$ (but not $|\beta_{12}|\,\tau_C$!) $\ll 1$ and $S_0 = 0$, equation~(\ref{eq:DelS}) reduces to
\begin{equation}
\label{eq:DelS1}
\Delta S_L / \Delta S_L^* \approx b_2(1 - \beta\tau_C/2)\;,
\end{equation}
where $\Delta S_L^*$ is the line flux density above the continuum in LTE. The term in the parentheses describes amplification of the radiation at the line frequencies due to the stimulation of line emission by the free-free continuum of the cloud.  Because $\tau_C \propto \nu^{-2.1}$ for the free-free continuum and because the maximum achievable values of $|\beta|$ increase with the decreasing line frequency (see Strelnitski \et\ 1996a), considerable amplifications with respect to $\Delta S_L^*$ are possible in this case only for low-frequency RRLs ($n\simgreat 100$).

{\it (2) Low opacities, strong continuum background} (Shaver 1978). The background continuum stimulates line transitions in this case, reducing equation~(\ref{eq:DelS}) to
\begin{equation}
\label{eq:DelS2}
\Delta S_L  \approx -\tau_L S_0\;.
\end{equation}
The line-to-continuum flux density ratio equals just $|\tau_L|$, and thus, if $\tau_L$ is not too small (say, $\simgreat 10^{-3}$), the line emission stimulated by the background source can make the line detectable on the continuum background. Shaver considered the synchrotron emission of a galactic nucleus as a continuum background. Then, again, only low-frequency lines can be amplified, because the brightness of the synchrotron emission drops with frequency.

{\it (3) $\tau_{net} \equiv (\tau_L + \tau_C)<-1$.} Since $\tau_C$ is always positive, $\tau_L$ should all the more be $<-1$ in this case. If $|\tau_{net}|$ is considerably greater than unity, equation~(\ref{eq:DelS}) becomes
\begin{equation}
\Delta S_L  \approx \biggl [B_{\nu}(T_e)\, \omega\, \eta + S_0\biggr ]  {\rm e}^{|\tau_{net}|}\;.
\label{eq:DelS3}
\end{equation}
Both the proper cloud emission (free-free continuum plus spontaneous line emission --- the first term in the brackets) and the continuous background (the second term) are amplified exponentially with the gain $|\tau_{net}|$. This is the ``true'' maser amplification, in contrast with the two previous cases --- of low-opacity and possibly even {\it positive} net value of the absorption coefficient, which Goldberg (1966) called ``partial masing.'' The ``true'' masing is more probable at high frequencies, where the free-free continuum is relatively weak. The first true, high-gain masers in mm hydrogen recombination lines (H$36\alpha$ to H$30\alpha$) were discovered three decades ago in the ionized disk/outflow of the peculiar emission-line star MWC 349 (Mart\'in-Pintado \et\ 1989). Later, it was found that masing continues into the submm domain and even into the far-IR (formally, already laser) domain, down to $n \approx 10$ at $\lambda \sim 50\,$mkm (Strelnitski \et\ 1996b).

May a first generation galaxy have H$^+$ regions that produce ``real'' maser lines, with the gain $|\tau_{net}| > 1$? For an estimate, we will use the model of Wood \& Loeb (2000) for a characteristic radial scale $R$, vertical scale $h$ and characteristic gas density $N_e(R)$ in a galaxy formed within a dark matter halo of mass $M_h$ at redshit $z$. In Strelnitski \et\ (1996a), based on extensive population calculations made by Storey \& Hummer (1995), the net coefficient of absorption for hydrogen $\alpha$ and $\beta$ lines in a broad interval of $n$ was calculated and presented in the tabular and graphical form. Using Fig.~8 from that paper, we find, for example, that in a galaxy forming at $z \approx 10$ in a halo of $M_h \sim 10^{10} M_{\odot}$, with $N_e(R) \sim 2\cdot 10^4$ cm$^{-3}$, and the vertical scale $h_0 \sim 10^{19}\,$cm at $r = R\sim 3\cdot 10^{20}\,$cm, the vertical path through the galaxy has an unsaturated maser gain $|\tau_{net}|\simgreat 1$ for $\alpha-$lines in a large range of $n \approx 40\,-\, 80$. For a galaxy forming in a halo of $M_h \sim 10^{11} M_{\odot}$ at this redshift, $N_e(R) \sim 10^5$ cm$^{-3}$, $h_0 \sim 5\cdot10^{19}\,$cm, and the unsaturated maser gain is $|\tau_{net}|\simgreat 10$ for $\alpha-$lines with $n \approx 30\,-\,70$. In these estimates, we supposed that all the gas on the line of sight is ionized. 

Note that the net absorption coefficients calculated in Strelnitski \et\ (1996a) were based on computer simulations for ordinary, contemporary H$^+$ regions. As it was argued in the previous sections, the conditions in the H$^+$ regions of the first-generation galaxies might be quite different and, in particular, these conditions might be conducive to higher population inversion in the lower-$n$ hydrogen $\alpha$-lines.

\subsection{Limitations Due to Saturation}

Unsaturated maser amplification is exponential. With the gain $|\tau_{net}| \approx 10$, the radiation brightness temperature would be raised by a factor $\approx {\rm e}^{10} \simgreat 10^4$. However, hydrogen recombination line masers are subject to relatively fast saturation (e.g. Strelnitski \et\ 1996a), which limits the output brightness temperature.  The condition of saturation is
\begin{equation}
J \approx I \frac{\Omega}{4 \pi} \simgreat J_s \equiv \frac{\Gamma}{B}\,,
\label{eq:J}
\end{equation}
where $J$ is the radiation intensity averaged over directions and line profile, $I$ is the intensity averaged over the line profile only, $\Omega$ is the solid angle of maser emission, $J_s$ is the ``saturation intensity,'' $\Gamma$ is the characteristic rate of decay for maser energy levels (a part of the pumping process), and $B$ is the Einstein $B$ coefficients for the maser transition. As long as $J << J_s$, maser amplification is exponential, with the gain $|\tau_{net}|$ [equation~(\ref{eq:DelS3})], but when the growing $J$ becomes $\simgreat J_s$, the amplification slows down and eventually becomes linear. 

Under  full saturation, each pumping cycle produces just one maser photon, which puts an upper limit on the photon luminosity of the source. The major pumping cycles in recombining hydrogen include spontaneous and collisional transitions. An adequate estimate for $\Gamma$ in equation~(\ref{eq:J}) is either the characteristic total rate of spontaneous decay of a maser level, $A_t$, or the characteristic total rate of its collisional decay, $C_t$, depending on which is faster.
\begin{figure}
\includegraphics[width=84mm]{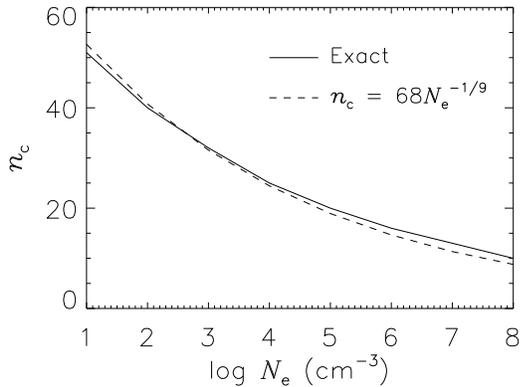}
%\vspace{3.5cm}
\caption{Dependence on electron density of the boundary level ($n_C$) for which the rates of spontaneous and collisional decay are approximately equal.}
\label{fig:12}
\end{figure}
Two curves in Figure~5.1 show the dependence on the electron density of the principal quantum number, $n_c$, of the level, for which the spontaneous rate and collisional rate for the transitions between two adjacent levels, $A_m$ and $C_m$, are approximately equal. Note that because of transitions to other levels, both $A_t$ and $C_t$ may be a few times greater than $A_m$ and $C_m$, respectively, but for rough estimates we aim at here, the ratios $A_t/C_t$ and $A_m/C_m$ can be considered approximately equal. The solid curve in Figure~5.1 was obtained by using the ``exact'' equations for $A_m$ (with the Gaunt factor $G = 1$) and $C_m$ (Van Regemorter 1962), while the broken curve is a simple empirical power-law  approximation of this function. For all the transitions below these curves, one can take $\Gamma \sim A_t$, and for those above the curves $\Gamma \sim C_t$.

The condition ``one maser photon per one pumping cycle'' under full saturation allows for an estimate of the upper limit for the photon luminosity of a maser source. The ratio of the maximum maser luminosity to spontaneous luminosity is $\rho \sim \Gamma/A_m$, which for transitions below $n_c$ is $\rho \sim A_t/A_m$ and for transitions above $n_c$ it is $\rho \sim C_t/A_m$. The ratio $A_t/A_m$ drops from $\approx 14$ for $n \approx 100$ to $\approx 2$ for $n\approx 5$. Thus one can expect an increase of luminosity due to masing/lasing by a factor of a few at best for all the lines below the $n_c$ curves in Fig.~5.1. Depending on the characteristic density of the galaxy, these are lines between, approximately, H5$\alpha$ and H30$\alpha$. On the other hand, the ratio $C_t/A_m \sim 5\cdot 10^{-17}n^9 N_e$ is a very steep function of $n$, which, in principle, can secure a very high rate of the maser level decay for high-$n$ levels and high gas densities. For example, in a galaxy formed within a $M_h = 10^{10} M_\odot$ halo at $z = 10$, with its characteristic density $N_e \sim 2\cdot 10^4\,$cm$^{-3}$, the $C_t/A_m$ ratio is $\simgreat 10^2$ for $\alpha$-lines with $n\simgreat 35$. At higher densities, the upper limit of saturated maser luminosity for high-$n$ transitions is proportionally higher.

We conclude that maser amplification can, in principle, increase the expected intensities of hydrogen recombination lines from the first-generation galaxies from a factor of a few, for the lines with $n \simless n_c$, to greater factors (up to orders of magnitude) for the lines with $n \simgreat n_c$. We emphasize, however, that this is only a $principal$ possibility, based on estimates of upper limits. In reality, the gains due to masing may be more modest.

\section{Concluding Remarks}

This study showed that the transitions between hydrogen Rydberg levels stimulated by the radiation of the ionizing star of an H$^+$ region may influence the distribution of the level populations significantly, and thus it is recommended to include these processes in the computations of hydrogen level populations both in contemporary H$^+$ regions and in those of the high-redshift galaxies. For the latter, an additional factor influencing the population distribution of higher levels ($n \ga 20$) is the warmer CMB.

Our computations of the hydrogen population distributions under the assumed physical conditions in the H$^+$ regions of high-redshift galaxies demonstrates a qualitative pattern similar to what had been obtained in several previous computations --- for contemporary H$^+$ regions: a sequence of inverted $\alpha$-transitions in a broad range of the principal quantum number $n$, the central $n$ value of the range decreasing when the electron density increases. The closeness of our results, obtained without the inclusion of the radiative processes induced by the star, to the results of the best previous studies confirms the reliability of the chosen method of solution.

When the processes induced by the star radiation, as well as by the radiation of the CMB (with the temperature corresponding to the redshift) are included, the distortion of the populations may be both ``positive'' and ``negative'' --- an increase or a decrease of the population inversion. A characteristic pattern for a high-$z$ (high $T_{\mathrm CMB}$) and high $T_*$ model is a decrease of inversion for the transitions with $n$ around the maximum inversion and an increase of inversion at lower and higher $n$'s (see Figs.~4.4 and 4.5).

In our calculations we ignored the transitions between $l$-sublevels induced by collisions with protons, therefore some more subtle effects noticed in previous studies, such as the appearance of ``over-cooled'' transitions (e.g. Strelnitski et al. 1996a) are not seen on our graphs. However, this simplification may be supported by recent studies which show that the effects of proton collisions are less significant than previously assumed (Vrinceanu \et\ 2012).

A potentially important result is the prediction of a possible population inversion of low$-n$ transitions, with $n \la 20$.  After redshift from $z \sim 10$, these transitions fall into the sub-millimeter domain. Although the effects of maser saturation should limit the possible maser (or rather laser) gains of these high-frequency transitions to a factor of 10 or so, this may be enough to make some of them detectable with sensitive modern radio interferometers, such as ALMA (see also Rule et al. 2013).

\section{Acknowledgments}

BP and KR were undergraduate research assistants at the Maria Mitchell Observatory while working on this project. They gratefully acknowledge the support by the NSF REU grant AST-0851892 and by the Nantucket Maria Mitchell Association.

\appendix
\section{The Rate of Photo-Excitation as a Function of Star Temperature}

The dominant process of interaction of hydrogen level $n$ with the radiation of the ionizing star is photo-excitation from the effective ground level, which is the level $n=2$ in the Menzel's Case B, more probable than Case A for the large and dense H$^+$ regions of the first galaxies. The derivative of the rate of this process with respect to the star's temperature is:
\begin{equation}
\label{eq:delT*}
\begin{split}
\frac{\partial}{\partial T_*}\left[N_2B_{2,n}J_{2,n}\right] &\propto \frac{\partial}{\partial T_*}\left(\frac{b_2}{\exp\left(h\nu/kT_*\right)-1}\right) \\
&= \left(\frac{\partial b_2}{\partial T_*}\right)\frac{1}{\exp\left(h\nu/kT_*\right)-1}\\
&+ b_2\frac{\partial}{\partial T_*}\left(\frac{1}{\exp\left(h\nu/kT_*\right)-1}\right),
\end{split}
\end{equation}
where $b_n$ is the ratio of the population of level $n$ to its LTE population, and $\nu$ is the frequency of the transition ($2\!\to\! n$). To determine the dependence of $b_2$ on $T_*$, we note that the population of the effective ground level is controlled by photo-ionization from this level and the recombination to \textit{all} the levels (because spontaneous cascade brings eventually most of the recombined atoms to the ground level). Ignoring the relatively insignificant stimulated recombination we have:
\begin{equation}
\label{eq:ionRecom}
N_2\int B_{2,\kappa}J_{2,\kappa}\>\rmn{d}\kappa = N_e^2\sum_{n=2}^\infty{\alpha_n^r}.
\end{equation}
Let
\begin{equation}
\int_1 = \int_{a/T_*}^\infty \! \frac{g^{II}}{x\left(e^x-1\right)}\,\mathrm{d}x
\label{eq:int1}
\end{equation}
and
\begin{equation}
\int_2 = \int_{I_n/kT_e}^\infty \! \frac{g^{II}e^{-x}}{x}\,\text{d}x,
\label{eq:int2}
\end{equation}
where $g^{II}$ is the bound-bound Kramers-Gaunt factor (Menzel \& Pekeris 1935),  $I_n = I_H/n^2$ is the energy of ionization from level $n$, $I_H$ is the energy of ionization from the ground level ($n=1$), and $a = I_H/4k$. From equation (34) in Burgess \& Summers (1976), (BS34),
\begin{equation}
\int \! B_{2,\kappa}J_{2,\kappa}\,\text{d}\kappa = \left(\frac{8\alpha^4c}{3\sqrt{3}\pi a_0}\right)\frac{W}{n^5}\int_1.
\label{eq:BS34}
\end{equation}
From (BS36),
\begin{equation}
\alpha_n^r = \frac{8}{n^3}\left(\frac{\pi a_0^2I_H}{kT_e}\right)^{3/2}\left(\frac{8\alpha^4c}{3\sqrt{3}\pi a_0}\right)e^{I_n/kT_e}\int_2,
\end{equation}
Plugging the above two equations into equation \eqref{eq:ionRecom},
\begin{multline}
%\begin{split}
N_2\left(\frac{8\alpha^4c}{3\sqrt{3}\pi a_0}\right)\frac{W}{n^5}\int_1 \\
= N_e^2\sum \frac{8}{n^3}\left(\frac{\pi a_0^2I_H}{kT_e}\right)^{3/2}\left(\frac{8\alpha^4c}{3\sqrt{3}\pi a_0}\right)e^{I_n/kT_e}\int_2,
%\end{split}
\label{eq:plug1}
\end{multline}
or, simplifying,
\begin{equation}
N_2\frac{W}{n^5}\int_1 = N_e^2\sum\frac{8}{n^3} \left(\frac{\pi a_0^2I_H}{kT_e}\right)^{3/2}e^{I_n/kT_e}\int_2.
\label{eq:plug2}
\end{equation}
From (BS53),
\begin{equation}
N_n = N_e^28\left(\frac{\pi a_0^2I_H}{kT_e}\right)^{3/2}n^2e^{I_n/kT_e}b_n.
\label{eq:BS53}
\end{equation}
Combining the above two equations and noting that in this case $n=2$,
\begin{multline}
N_e^28\left(\frac{\pi a_0^2I_H}{kT_e}\right)^{3/2}2^2e^{I_2/kT_e}b_2\frac{W}{2^5}\int_1 \\
= N_e^2\sum8\left(\frac{\pi a_0^2I_H}{kT_e}\right)^{3/2}\frac{1}{i^3}e^{I_n/kT_e}\int_2.
\label{eq:comb}
\end{multline}
Simplifying again gives
\begin{equation}
\frac{W}{8}e^{I_2/kT_e}b_2 = \sum\frac{1}{n^3}e^{I_n/kT_e}\int_2.
\label{eq:combSimp}
\end{equation}
And then:
\begin{equation}
b_2\int_1 = \left(\frac{8}{W}\right)e^{-I_2/kT_e}\sum\frac{1}{n^3}e^{I_n/kT_e}\int_2,
\label{eq:2.5}
\end{equation}
or,
\begin{multline}
%\begin{split}
b_2\int_{a/T_*}^\infty {\frac{g^{II}}{x\left(e^x-1\right)}}\>\rmn{d}x = \left(\frac{8}{W}\right)e^{-a/T_e}\\
\times \sum_{n = 2}^\infty \left(\frac{1}{n^3}e^{I_n/kT_e}\int_{I_n/kT_e}^\infty \frac{g^{II}e^{-x}}{x}\>\rmn{d}x\right)\;.
%\end{split}
\end{multline}
Since the right-hand side of this equation is independent of $T_*$, so too must be the left-hand side.  Therefore, its derivative with respect to $T_*$ must vanish, \textit{i.e.},
\begin{equation}
\begin{split}
\left(\frac{\partial b_2}{\partial T_*}\right)\int_{a/T_*}^\infty &{\frac{g^{II}}{x\left(e^x-1\right)}}\>\rmn{d}x\\
&+ b_2\frac{\partial}{\partial T_*}\left(\int_{a/T_*}^\infty {\frac{g^{II}}{x\left(e^x-1\right)}}\>\rmn{d}x\right) = 0.
\end{split}
\end{equation}
Therefore,
\begin{equation}
\label{eq:delTbn}
\frac{\partial b_2}{\partial T_*} = -\frac{b_2g^{II}}{T_*\left(e^{a/T_*}-1\right)}\left(\int_{a/T_*}^\infty {\frac{g^{II}}{x\left(e^x-1\right)}}\>\rmn{d}x\right)^{-1}.
\end{equation}
Noting that $h\nu/k \approx a$ for $n \gg 2$,
\begin{equation}
\label{eq:delTPEX}
\frac{\partial}{\partial T_*}\left(\frac{1}{\exp\left(h\nu/kT_*\right)-1}\right) = \frac{ae^{a/T_*}}{T_*^2\left(e^{a/T_*}-1\right)^2}.
\end{equation}
Combining equations (\ref{eq:delT*}), (\ref{eq:delTbn}), and (\ref{eq:delTPEX}), we finally get:
\begin{equation}
\begin{split}
\frac{\partial}{\partial T_*}&\left[N_2B_{2,n}J_{2,n}\right] \propto \frac{b_2}{T_*^2}\left(e^{a/T_*}-1\right)^{-2}\\
&\times \left[ae^{a/T_*} - T_*g^{II}\left(\int_{a/T_*}^\infty {\frac{g^{II}}{x\left(e^x-1\right)}}\>\rmn{d}x\right)^{-1}\right].
\end{split}
\label{eq:negRate}
\end{equation}
Since $a$, $T_*$, and $b_2$ are all positive, the sign of equation (\ref{eq:negRate}) is determined solely by the sign of the bracketed expression. It can be easily verified that for all the relevant values of $T_*$ (\textit{e.g.} for all $T_*\la10^{6}\,$K), this expression is negative, and therefore,
\begin{equation}
\frac{\partial}{\partial T_*}\left[N_2B_{2,n}J_{2,n}\right] < 0.
\label{eq:B2Tdep}
\end{equation}
We conclude that the rate of photo-excitation from level $n = 2$ decreases with increasing stellar temperature. This conclusion is confirmed by our calculations of $\xi_1$ and $\xi_2$ [equation \eqref{eq:PEXratios}] for various values of $T_*$. One can expect, therefore, that the cooling effect on population distribution also decreases with the increasing stellar temperature; and thus, the calculated degree of inversion will be greater in the presence of a hotter star.

However, equation (\ref{eq:B2Tdep}) is derived on the assumption that the major depopulation process for the effective ground level is photoionization by the stellar radiation.  At sufficiently low temperatures, this assumption is no longer valid.  At star temperatures $T_* \la 1\times 10^4\,$K and in the range of densities considered here, the main process that depletes the population of the effective ground level is collisional excitation to higher levels. For stars below this temperature, the sign of equation \eqref{eq:B2Tdep} switches and any further decrease of $T_*$ is accompanied by a decrease in the rate of depopulation of the effective ground level to higher levels.  This change results in a decrease in the cooling effect of the stellar radiation, and the solution for the relative level populations (the $\beta$ coefficient; see sections 3 and 4) tends again to the ``no star'' solution (see Fig. \ref{fig:11}).

\section{The Rate of Photo-Excitation as a Function of Dilution Factor}

 In order to see whether and how the influence of star's radiation on the distribution of populations depends on the dilution factor, we analyze the derivative
\begin{equation}
\label{eq:delW}
\begin{split}
\frac{\partial}{\partial W}\left[N_2B_{2,n}J_{2,n}\right] = \left(\frac{\partial N_2}{\partial W}\right)&B_{2,n}J_{2,n}\\
&+ N_2\frac{\partial}{\partial W}\left(B_{2,n}J_{2,n}\right).
\end{split}
\end{equation}
Using equation (\ref{eq:ionRecom}) again, we see that
\begin{equation}
\frac{\partial N_2}{\partial W}\int B_{2,\kappa}J_{2,\kappa}\>\rmn{d}\kappa + N_2\frac{\partial}{\partial W}\left(\int B_{2,\kappa}J_{2,\kappa}\>\rmn{d}\kappa\right) = 0.
\end{equation}
Since, assuming $W\neq0$,
\begin{equation}
\frac{\partial}{\partial W}\int B_{n,\kappa}J_{n,\kappa}\>\rmn{d}\kappa = \frac{1}{W}\int B_{n,\kappa}J_{n,\kappa}\>\rmn{d}\kappa,
\end{equation}
we have,
\begin{equation}
\label{eq:delWN}
\frac{\partial N_2}{\partial W} = -\frac{N_2}{W}.
\end{equation}
It is readily seen that
\begin{equation}
\label{eq:delWPEX}
\frac{\partial}{\partial W}\left(B_{2,n}J_{2,n}\right) = \frac{B_{2,n}J_{2,n}}{W}.
\end{equation}
Putting together equations (\ref{eq:delW}), (\ref{eq:delWN}), and (\ref{eq:delWPEX}),
\begin{equation}
\frac{\partial}{\partial W}\left[N_2B_{2,n}J_{2,n}\right] = 0\;.
\end{equation}
This result suggests that the change of the dilution factor across the H$^+$ region should not have any significant impact on the distribution of the level populations.  Of course, this derivation also relies on the assumption that the depopulation of level $n=2$ is controlled by photodissociation.  As we discussed above, this assumption is adequate for all the range of the star temperatures of interest here ($1\times 10^4 \la T_* \la 1\times 10^5\,$K). 

\label{lastpage}

\end{document}